\documentstyle[preprint,tighten,aps,floats,epsfig]{revtex}

\begin{document}
\draft
\vskip 2cm

\title{The Lattice Free Energy with Overlap Fermions: A Two-Loop Result}

\author{A. Athenodorou and H. Panagopoulos}
\address{Department of Physics, University of Cyprus, P.O. Box 20537,
Lefkosia CY-1678, Cyprus \\
{\it email: }{\tt ph00aa1@ucy.ac.cy, haris@ucy.ac.cy}}
\vskip 3mm

\date{\today}

\maketitle

\begin{abstract}

We calculate the 2-loop partition function of QCD on the lattice,
using the Wilson formulation for gluons and the overlap-Dirac operator
for fermions. Direct by-products of our result are the 2-loop free
energy and average plaquette.

Our calculation serves also as a prototype for further higher loop
calculations in the overlap formalism.

We present our results as a function of a free parameter $M_0$
entering the overlap action; the dependence on the number of colors
$N$ and fermionic flavors $N_f$ is shown explicitly. 


\end{abstract}

\newpage


\section{Introduction}
\label{introduction}

In recent years there has been great progress toward the implementation of chirality preserving
regularizations of gauge theories coupled to fermions on the lattice.
In the two most widely studied approaches, overlap fermions and domain wall fermions (for recent
reviews, see~\cite{Chiu,CW}), numerical simulations are already producing promising results.

Alongside with numerical simulations, a number of perturbative
calculations have also been carried out~\cite{AFPV,Capitani,Schierholz,YAK}, 
which allow for a meaningful comparison between lattice results and continuum quantities.
However, unlike the more conventional fermionic actions, the
complexity of the overlap and domain wall actions has hampered the scope
of perturbative results, and has kept the order of perturbative calculations down to 1 loop.

In this paper we perform a first 2-loop calculation using the overlap-Dirac
fermionic action. The quantity we compute is the expectation value of
the plaquette, in $SU(N)$ gauge theory coupled to $N_f$ massless fermionic species.
Aside from serving as a longtime testing ground for lattice
perturbation theory, the plaquette operator has appeared in a variety
of lattice investigations. One of its earliest uses has been in
attempts to extract the quadratic gluonic condensate~\cite{DR,CDG}.
Another use regards the definition of an effective coupling
constant~\cite{MPP,AFP}, in an effort to improve the scaling properties of observables;
this definition leads also to a modification of asymptotic scaling,
which involves higher-loop coefficients of the plaquette. Finally, the
plaquette is a standard ingredient in studies of the quark-antiquark
potential (see, e.g.,~\cite{BB} and references therein).

In standard notation, we write the gluonic and fermionic parts of the action as follows: 
\begin{eqnarray}
  S &=& S_G + S_F , \nonumber \\
 S_G &=& \beta \sum_\Box E_G(\Box ) , \qquad
 S_F = \sum_{x,y} E_F(x,y)
\label{totalaction}
\end{eqnarray}
with
\begin{eqnarray}
\sum_\Box E_G(\Box ) &=& \sum_\Box \left(1 - {1\over N} \hbox{Re Tr} \left(\Box \right)\right) \nonumber \\
&=& {1\over 2}\sum_{x}\sum_{\mu\ne\nu} \left(1 - {1\over N} \hbox{Re Tr} 
\left(U_{x,x+\hat\mu}\,U_{x+\hat\mu,x+\hat\mu+\hat\nu}\,
U_{x+\hat\mu+\hat\nu,x+\hat\nu}\, U_{x+\hat\nu,x}\right)\right), \\
 E_F(x,y) &=& \sum_{i=1}^{N_f} a^4 \bar\psi_i(x)\,D_{\rm N}(x,y)\,\psi_i(y) \label{actionF}
\end{eqnarray}
As usual, $\beta\equiv 2N/g_0^2$ is the lattice coupling and $a$ is the lattice spacing; 
$\Box$~stands for the plaquette and $U_{x,x+\mu}$ for the link
variable joining neighboring sites in the $\hat\mu$ direction: $x$ and $x +\hat\mu$. 
The Neuberger-Dirac fermionic operator $D_N$ depends on the link variables; its precise form will be presented below.

The average value of the action density, $S/V$, is directly related to
the average plaquette; in particular, for the gluonic part we have: 
\begin{equation}
\langle S_G/V \rangle = 6 \,\beta\,\langle E_G(\Box)\rangle, \qquad \langle E_G(\Box)\rangle = 1-{1\over N}\,\langle {\rm
Tr}(\Box)\rangle.
\end{equation} 
As for $\langle S_F/V\rangle$, it is trivial in any action which is
bilinear in the fermion fields~\cite{AFP}; this can be seen by rescaling $S_F$ in the fermionic partition function by a factor $\epsilon$ :
\begin{equation}
 Z^F(\epsilon) \equiv \int [{\cal{D}}\bar\psi_i
               {\cal{D}}\psi_i] \exp\left(- \epsilon S_F \right) = \epsilon^{4 V N N_f} 
               Z^F(\epsilon =1) \label{Zf}
\end{equation}
There follows:
\begin{equation}
 \langle \sum_y E_F(x,y) \rangle = -  {\partial \over {\partial \epsilon}}
                       \bigl({{\ln Z^F(\epsilon)} \over V}\bigr)_{\epsilon =1} = - 4 N N_f
\label{ef}
\end{equation}

We will thus calculate $\langle E_G(\Box)\rangle$ in perturbation theory:
\begin{equation}
\langle E_G \rangle = c_1 \, g_0^2 + c_2 \, g_0^4 + c_3 \, g_0^6 + \cdots
\label{expansion}
\end{equation}
The $n$-loop coefficient can be written as $c_n = c^G_n + c^F_n$ where $c^G_n$ is the contribution of diagrams without fermion loops and $c_n^F$ comes from diagrams containing fermions. The coefficients $c^G_n$ have been known for some time up to 3
loops~\cite{ACFP,AFP}; more recently, some estimates for higher loops have been produced in 4 dimensions using stochastic
methods~\cite{DiRenzo}, in the absence of fermions. The coefficients $c_n^F$ are also known to 3 loops for Wilson fermions~\cite{AFP} and clover
fermions~\cite{PT}. The task at hand, $c_n^F$ for overlap fermions, presents many more complications, and we will be computing $c_2^F$ in the present work.

The calculation of $c_n$ proceeds most conveniently by computing first the free energy $-(\ln Z)/V$, where $Z$ is the full partition function
\begin{equation}
Z \equiv \int [{\cal D}U {\cal D}\bar\psi_i {\cal D}\psi_i] \exp(-S) .
\label{Z}
\end{equation}
The average of $E_G$ is then extracted as follows
\begin{equation}
\langle E_G \rangle = - {1 \over 6}\, {\partial \over {\partial \beta}}\, \left( {\ln Z \over V} \right) .
\label{e}
\end{equation}
In particular, the perturbative expansion of $(\ln Z)/V$ :
\begin{equation}
(\ln Z)/V = {\rm constant} + \left( -{3 (N^2-1)\over 2}\,\ln\beta + {d_1\over\beta} + {d_2\over\beta^2} + \cdots \right)
\end{equation}
leads immediately to the relations: $c_2= d_1/(24N^2)$, $c_3= d_2/(24N^3)$.

A total of 8 Feynman diagrams contribute to the present calculation, up to 2-loop order; these are shown in Figure 1.

All the algebra necessary to carry out this calculation was done with our computer package written in Mathematica; to this end, certain
extensions of the package were needed, in order to incorporate the numerical integration of 2-loop expressions involving overlap fermions.
The first 6 diagrams in Figure 1 give the purely gluonic contributions to the free energy: $c_1^G$, $c_2^G$, which are well known (see,
e.g.,~\cite{AFP}). Their contribution is presented here for the sake of completeness:
\begin{eqnarray}
 c_1^G &=& {{N^2 - 1} \over 8 \;N} , \nonumber \\
 c_2^G &=& \left( N^2 - 1 \right) \left(0.0051069297 - {1 \over {128 \; N^2}} \right)
\label{cg}
\end{eqnarray}

\begin{center}
\psfig{figure=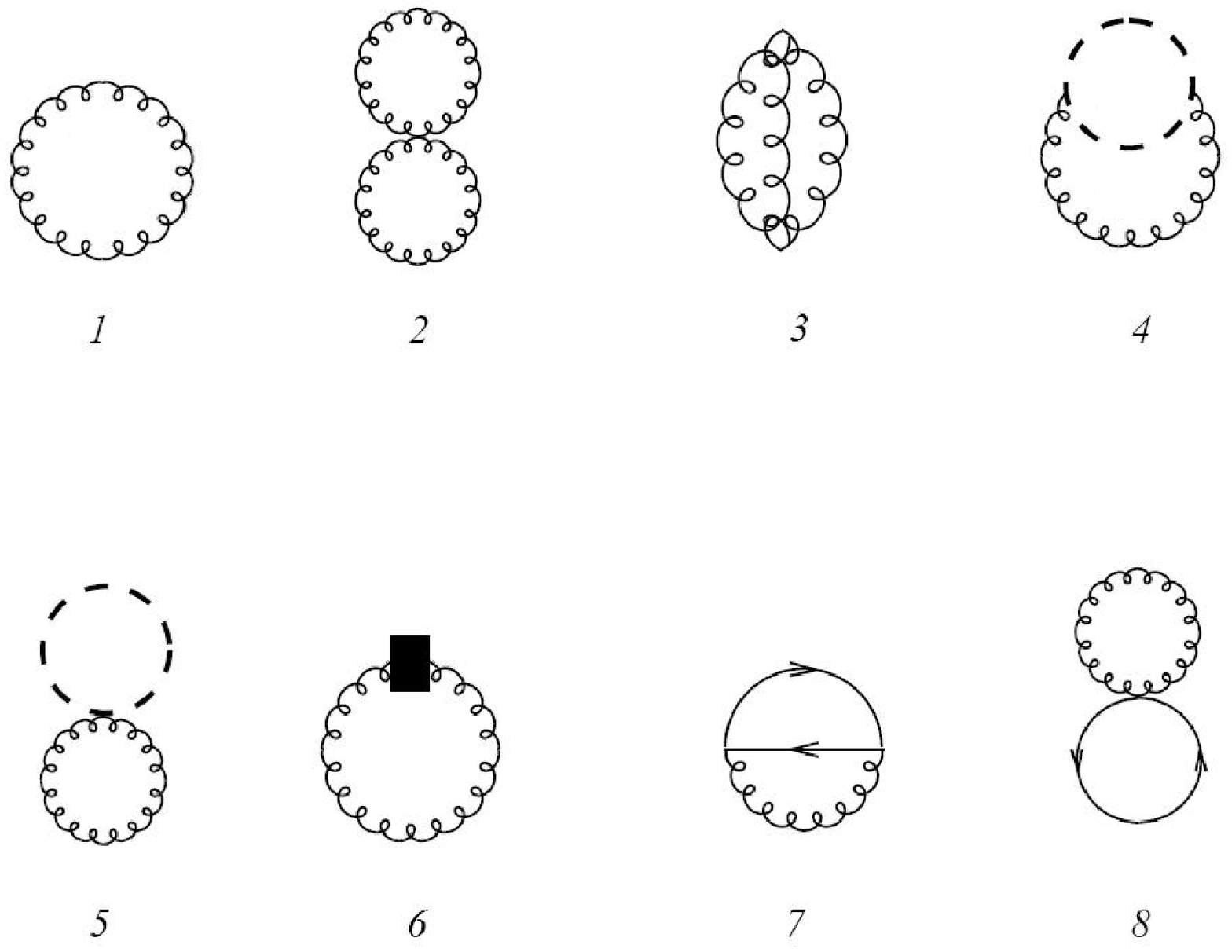,height=12truecm}
\end{center}
\begin{center}
{\footnotesize Fig. 1 Feynman diagrams contributing to the free energy up to 2 loops. Solid, wavy, dashed lines denote fermions, gluons and ghosts,
respectively. A solid square stands for a contribution from the measure part of the action.}
\end{center}
\medskip

Diagrams 7 and 8 are much more cumbersome to calculate; they lead to the value of $c_2^F$, whose dependence on $N$ and $N_f$ can be made explicit:
\begin{equation}
c_2^F = {N^2 -1 \over 12\, N} \, N_f \, h_2\, ,
\end{equation}
where $h_2$ is the quantity we set out to calculate; it depends only on a parameter $M_0$ which appears in the overlap action, to which we now turn.

The Neuberger-Dirac operator $D_N$ appearing in the fermionic part of the action, Eq.(\ref{actionF}), has the form~\cite{Neuberger-98}:
\begin{eqnarray}
D_{\rm N} &=& {1\over a} \,M_0 \left(  1 + X {1\over \sqrt{X^\dagger X}} \right),\\
X &=& a \, D_{\rm W} - M_0, \label{Nop}
\end{eqnarray}
where $D_{\rm W}$ is the Wilson-Dirac operator (with the Wilson parameter $r$ set to its standard value, $r=1$)
\begin{equation}
D_{\rm W} = {1\over 2} \left[ \gamma_\mu \left( \nabla_\mu^*+\nabla_\mu\right)
 - a\nabla_\mu^*\nabla_\mu \right], \qquad
\nabla_\mu\psi(x) = {1\over a} \left[ U(x,\mu) \psi(x + a\hat{\mu})- \psi(x)\right],
\end{equation}
$M_0$ is a real parameter, whose value may be chosen at will; in order for this action to describe one massless fermion, with no doubling
problem, $M_0$ must lie in the range $0< M_0 < 2$. Nonperturbatively one expects
$-m_c < M_0 < 2$, where $m_c<0$ is the critical mass associated with the Wilson-Dirac operator.

Despite the non-ultralocal nature of the square root $[X^\dagger X]^{1/2}$, the operator $D_N$ is nevertheless local under reasonable
assumptions regarding the gauge configuration~\cite{Luscher}. In particular, the expansion of $D_N$ to all orders in perturbation
theory, while very elaborate, poses no conceptual obstacle and can be performed starting from an integral representation of the square root~\cite{KY}.

The expressions for the overlap propagator and vertices are presented in the Appendix. It is worth noting that, by the nature
of the overlap operator, the fermion vertex with two gluons contains two distinct parts: $\Sigma_2 = \Sigma_{2\alpha} + \Sigma_{2\beta}$ (see
Appendix A); the first of these is pointlike, while the second involves integration over an internal four-momentum, as shown
in Fig. 5. Consequently, diagram 8 splits into parts 8a, 8b; diagram 8b has the same topology as diagram 7.

The numerical integration over loop momenta was carried out for a wide range of values for $0<M_0<2$ on finite
lattices of size $L\le 128$, with subsequent extrapolation to infinite size. The systematic error resulting from the extrapolation is
typically small enough to leave the first six significant digits intact. A very slight reduction in precision is noticed as $M_0 \to 2$, where
the pole structure of the fermionic propagator is expected to change. 

Our results, as a function of $M_0$, are presented in Figure 2 for each fermionic diagram separately; their sum is shown in Figure 3 and
listed in Table I. It is worth noting that the results are very smooth functions of $M_0$, despite the potential singularities of the integrands at $M_0=0$ and $M_0=2$.

To facilitate comparison, we provide below the values of individual diagrams for a particular value of $M_0$:
\begin{eqnarray}
&&h_2^7(M_0{=}1) =   -0.02013079(1), \qquad h_2^{8\alpha}(M_0{=}1) = -0.0099775548(1), \nonumber \\
&&h_2^{8\beta}(M_0{=}1)  = 0.010764939(1)
\end{eqnarray}

\begin{center}
\psfig{figure=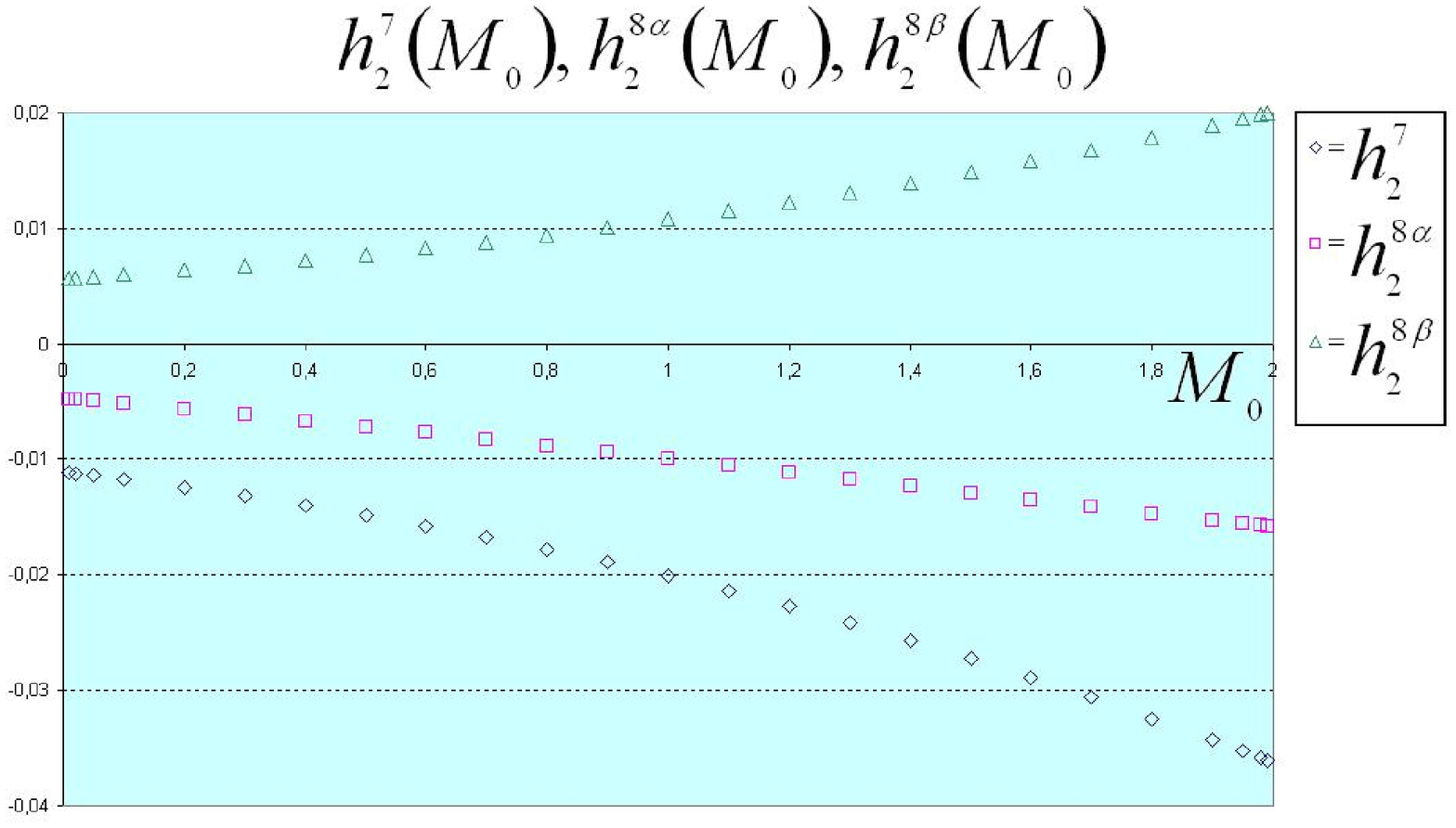,height=10truecm}\\
{\footnotesize Fig. 2 Numerical results for each diagram involving fermions, as a 
function of the parameter $M_0$. The values denoted by $8\alpha$ and $8\beta$
correspond to the two contributions to diagram 8, coming from vertices $\Sigma_{2\alpha}$ and $\Sigma_{2\beta}$, respectively.}
\end{center}

\begin{center}
\psfig{figure=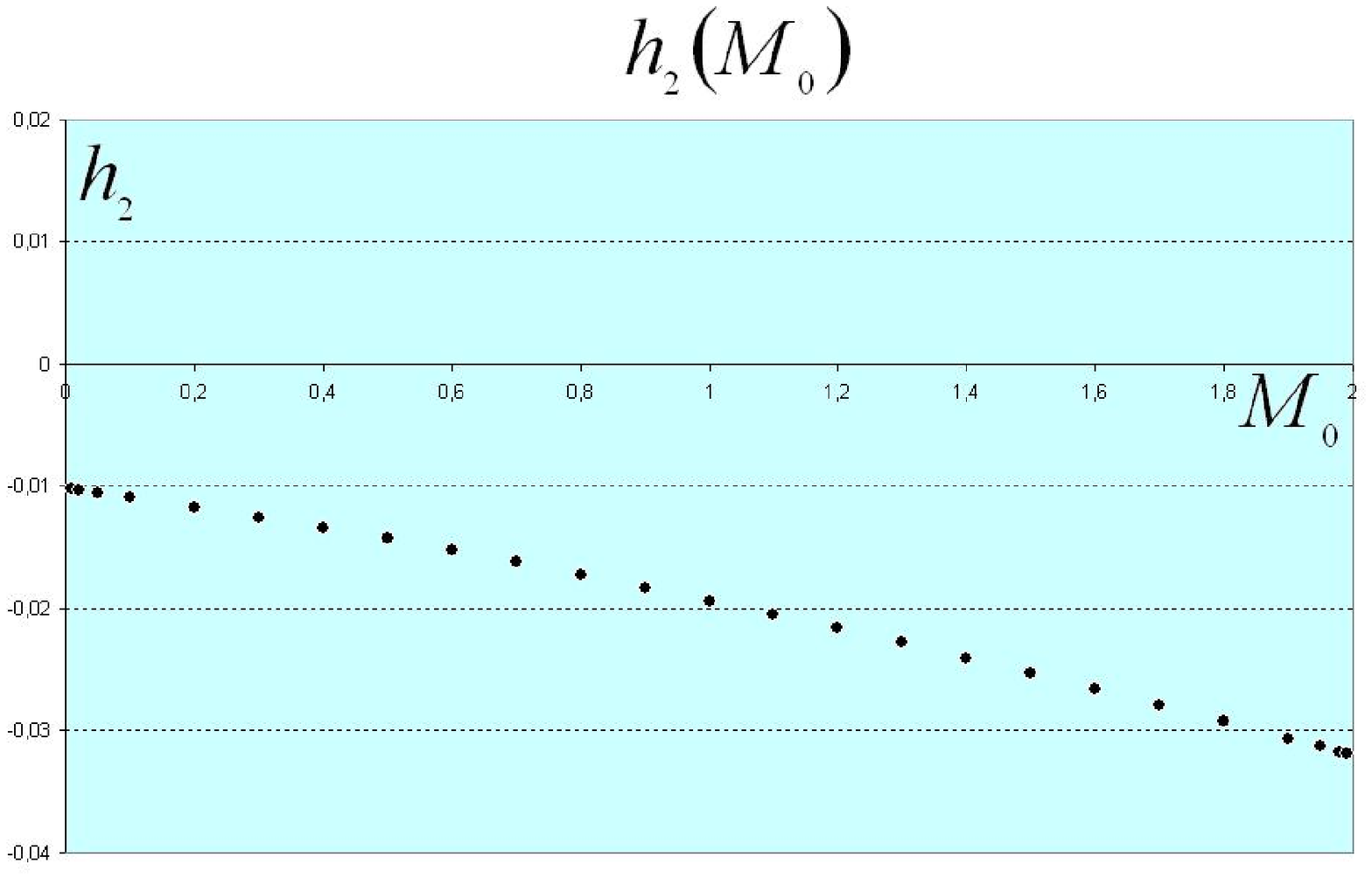,height=10truecm}\\
{\footnotesize Fig. 3 The total contribution of fermionic diagrams to $h_2$, as a function of the parameter $M_0$\,.}
\end{center}
\medskip

For ease of reference, we list below the total value of $\langle E_G(\Box)\rangle$ for $N=3$ and $N_f=0,2,3$, with some typical values of
the action parameters ($\kappa$ is the hopping parameter of the Wilson action):
\begin{equation}
\begin{array}{llllll}
\langle E_G(\Box)\rangle &=& g_0^2/3 \ + &g_0^4 \ \cdot &
  0.033910993\qquad&(N_f=0) \\
&=& g_0^2/3 \ + &g_0^4 \ \cdot &0.0234689 &(N_f=2\ \ {\rm Wilson},\ r=1,\ \kappa=0.156) \\
&=& g_0^2/3 \ + &g_0^4 \ \cdot &0.0253139 &(N_f=2\ \ {\rm overlap},\ M_0=1.0) \\
&=& g_0^2/3 \ + &g_0^4 \ \cdot &0.0226752 &(N_f=2\ \ {\rm overlap},\ M_0=1.5) \\
&=& g_0^2/3 \ + &g_0^4 \ \cdot &0.0182479 &(N_f=3\ \ {\rm Wilson},\ r=1,\ \kappa=0.156) \\
&=& g_0^2/3 \ + &g_0^4 \ \cdot &0.0210154 &(N_f=3\ \ {\rm overlap},\ M_0=1.0) \\
&=& g_0^2/3 \ + &g_0^4 \ \cdot &0.0170573 &(N_f=3\ \ {\rm overlap},\ M_0=1.5) \\
\end{array}
\end{equation}                           

The extension of the above calculation in order to encompass Wilson loops of larger size, as may be required for computing the
quark-antiquark potential, presents no additional difficulties. Further two-loop calculations with the overlap action
involve fermionic vertices with 3 and 4 gluons; generating these vertices is conceptually straightforward, though technically it is
quite involved. We are presently working on these extensions, and expect to provide results in a future report.

\appendix

\section{}
\label{appa}

We present here the expressions for the overlap propagator and for the fermionic vertices containing up to two gluons,
following Refs.~\cite{KY,AFPV}. This Appendix serves for completeness, and also in order to correct some typographical errors
(which had no consequence on previous results).

Let us first write down the weak coupling expansion of the Wilson-Dirac operator $D_{\rm W}$. This will be useful for
constructing the relevant vertices of $D_{\rm N}\,$. We write
\begin{eqnarray}
X(q,p) &\equiv& a^4 \, \sum_{x,y} e^{-i q \cdot x \, a + i p \cdot y \,
a} \, (D_{\rm W}(x,y) - {M_0\over a}\, \delta_{x,y}) \nonumber \\
& =& X_0(p)(2\pi)^4\delta^4(q-p) + X_1(q,p) + X_2(q,p) + {\cal O}(g_0^3),
\label{aqp}
\end{eqnarray}
\begin{eqnarray}
{\rm where: }\qquad X_0(p)&=& {i\over a} \sum_\mu \gamma_\mu \sin a p_\mu + {1\over a} \sum_\mu (1-\cos a p_\mu) - {1\over a} M_0,
\label{a0}\\
X_1(q,p) &=& g_0 \int d^4 k \delta(q-p-k) A_\mu(k) V_{1,\mu}(p+k/2),\label{a1}\\
V_{1,\mu}(q) &=& i\gamma_\mu \cos aq_\mu + \sin a q_\mu,\nonumber
\end{eqnarray}
\begin{eqnarray}
X_2(q,p) &=& {g_0^2\over 2} \int {d^4 k_1 \, d^4 k_2\over (2\pi)^4}
 \delta(q-p-k_1-k_2) A_\mu(k_1)A_\mu(k_2) V_{2,\mu}(p+k_1/2+k_2/2),\label{a2}
\\
V_{2,\mu}(q) &=& -i\gamma_\mu a\sin aq_\mu + a \cos a q_\mu.\nonumber
\end{eqnarray}

The Fourier transform of the Neuberger-Dirac operator takes the form
\begin{equation}
{1 \over M_0} D_{\rm N}(q,p) = D_0(p) (2\pi)^4\delta^4(q-p) + \Sigma(q,p).
\end{equation}
The factor $1/M_0$ in the above equation is inconsequential, since it
can be absorbed in the definition of $\psi(x)$. $D_0(p)$ is the tree level inverse propagator: 
\begin{equation}
D_0^{-1}(p) = {-i\sum_\mu \gamma_\mu \sin ap_\mu \over 2 \left[ \omega(p) + b(p)\right] } + {a\over 2},
\label{d0}
\end{equation}
\begin{eqnarray}
{\rm where:}\qquad \omega(p) &=& {1\over a} \left( \sum_\mu \sin^2 ap_\mu + \bigl[ \sum_\mu (1-\cos ap_\mu ) - M_0 \bigr]^2 \right)^{1/2}, \\
b(p)&=& {1\over a} \sum_\mu (1-\cos ap_\mu) - {1\over a} M_0.
\end{eqnarray}
The function $\Sigma(q,p)$ can be expanded in powers of $g_0$ as
\begin{eqnarray}
a\Sigma(q,p) = && {1\over \omega(q) + \omega(p)}
\left[X_1(q,p) - {1\over \omega(q)\omega(p)} X_0(q) X^\dagger_1(q,p) X_0(p)\right] \nonumber\\
&&+ {1\over \omega(q) + \omega(p)}
\left[X_2(q,p) - {1\over \omega(q)\omega(p)} X_0(q) X^\dagger_2(q,p) X_0(p)\right] \nonumber\\
&&+\int {d^4 k\over (2\pi)^4}
{1\over \omega(q) + \omega(p)}{1\over \omega(q) + \omega(k)}{1\over \omega(k) + \omega(p)}\times \nonumber \\
&&\;\;\Biggl[-X_0(q)X_1^\dagger(q,k)X_1(k,p) 
 -X_1(q,k)X_0^\dagger(k)X_1(k,p) - X_1(q,k)X_1^\dagger(k,p)X_0(p) \nonumber \\
&&\;\;+{\omega(q)+\omega(k)+\omega(p)\over \omega(q)\omega(k)\omega(p)}
X_0(q)X_1^\dagger(q,k)X_0(k)X_1^\dagger(k,p)X_0(p)\Biggr] + {\cal O}(g_0^3)
\label{vertices}
\end{eqnarray}
From $\Sigma(q,p)$ one can read off the vertices containing up to two gluons. The vertex with one gluon, $a\Sigma_1(q,p)$ corresponds to the
first line of Eq.(\ref{vertices}).

\begin{center}
\begin{minipage}{0.4\linewidth}
\begin{center}
\psfig{figure=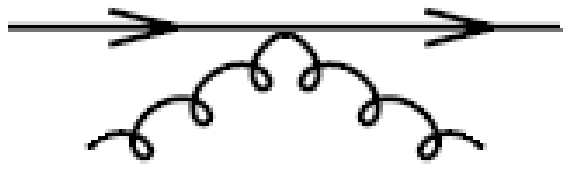,height= 2truecm}\\
\end{center}
\vskip-2.7cm \hskip1.7cm $\vec p$\hskip2.8cm $\vec q$\vskip2.7cm
\vskip-0.5cm
\hskip0.8cm{\footnotesize Fig. 4 Pointlike part of the vertex.\\
\phantom{Some more space here...}}
\end{minipage} \hskip0.1\textwidth
\begin{minipage}{0.4\linewidth}
\begin{center}
\psfig{figure=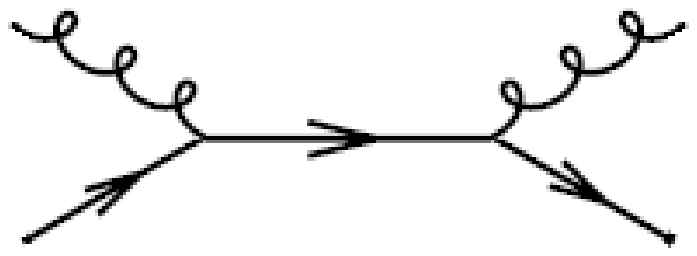,height= 2truecm}\\
\end{center}
\vskip-1cm \hskip0.8cm $\vec p$\hskip4.5cm $\vec q$ \vskip1cm
\vskip-2.8cm \hskip3.3cm $\vec k$ \vskip2.8cm
\vskip-1cm
{\footnotesize Fig. 5 Vertex part with intermediate momentum
integration}
\end{minipage}
\end{center}
\medskip

The fermionic vertex with two gluons splits into a pointlike part $a\Sigma_{2\alpha}$ (Fig. 4, second line of Eq.(\ref{vertices})) plus
a part involving an internal four-momentum $k$, which must be
integrated over, $a\Sigma_{2\beta}$ (Fig. 5, last 3 lines of
Eq.(\ref{vertices})).

\begin{table}[ht]
\begin{center}
\begin{minipage}{7cm}
\caption{Two-loop fermionic contribution $h_2$, for various values of
$M_0$,\ \  $0<M_0<2$. 
\label{tab1}}
\begin{tabular}{r@{}lr@{}l}
\multicolumn{2}{c}{$M_0$}&
\multicolumn{2}{c}{$h_2$} \\
\tableline \hline
0&.01   & -0&.01021118(5) \\
0&.05   & -0&.01050736(5) \\
0&.1    & -0&.01088833(5) \\
0&.2    & -0&.01168252(5) \\
0&.4    & -0&.01338658(3) \\
0&.6    & -0&.01523365(1) \\
0&.8    & -0&.01721907(1) \\
1&.     & -0&.01934341(1) \\
1&.2    & -0&.02160970(1) \\
1&.4    & -0&.02402085(3) \\
1&.6    & -0&.02657516(3) \\
1&.8    & -0&.02925538(3) \\
1&.9    & -0&.03062501(5) \\
1&.95   & -0&.03130974(5) \\
1&.99   & -0&.0318533(3)  \\
\end{tabular}
\end{minipage}
\end{center}
\end{table}


\end{document}